\documentclass[nocopyrightspace,preprint,nonatbib,10pt]{sigplanconf}
\pdfoutput=1

\let\bibfont\relax

\expandafter\let\csname ver@natbib.sty\endcsname\relax

\usepackage[backend=bibtex,style=alphabetic]{biblatex}
\renewcommand*{\bibfont}{\small}

\title{Building Code with Dynamic Staging
}

\newcommand{\inst}[1]{\textsuperscript{#1}}
\authorinfo{Piotr Danilewski\inst{1,2,3} \and Philipp Slusallek\inst{1,2,4}}
{\inst{1}Saarland University, Germany
\and
\inst{2}Intel Visual Computing Institute, Germany
\and\\
\inst{3}Theoretical Computer Science, Jagiellonian University, Poland
\and\\
\inst{4}Deutsches Forschungszentrum f\"{u}r K\"{u}nstliche Intelligenz, Germany}
{piotr.danilewski@tcs.uj.edu.pl\hspace{5mm}slusallek@cg.uni-saarland.de\vspace{-.5cm}}
\usepackage{etex}
\usepackage[utf8]{inputenc}
\usepackage[T1]{fontenc}

\usepackage{amsfonts}

\usepackage{microtype}

\usepackage{float}
\usepackage{acronym}
\usepackage{amsthm} 
\usepackage{amssymb}
\usepackage{amsmath}
\usepackage{mathtools}
\usepackage{multirow}
\usepackage{listings}
\usepackage{newfloat}
\usepackage{relsize}
\usepackage{verbatim}
\usepackage[labelfont=bf]{caption}
\usepackage[labelfont=bf,subrefformat=parens]{subcaption}
\usepackage{tikz}
\usepackage{ctable}
\usetikzlibrary{arrows,decorations,backgrounds,positioning,fit,petri}
\usetikzlibrary{shapes.multipart}
\usetikzlibrary{fadings}
\usetikzlibrary{patterns}
\usetikzlibrary{decorations.pathmorphing}

\ifx\pdfoutput\undefined
    \usepackage[final]{graphicx}
    \usepackage{hyperref}
\else
    \usepackage[pdftex]{hyperref}
    \hypersetup{pdftex,linkcolor=blue,citecolor=magenta,urlcolor=red,colorlinks=true}
\fi

\usepackage{refcount}
\newcounter{fncntr}
\newcommand{\footmark}[1]{\footnotemark[\getrefnumber{#1}]}
\newcommand{\foottext}[2]{\refstepcounter{fncntr}\label{#1}\footnotetext[\getrefnumber{#1}]{#2}}
\input{template/tikzshapes.tex}
\definecolor{darkgreen}{rgb}{0.2,0.7,0.1}
\lstdefinelanguage{CUDA}
[ISO]{C++}
{
morekeywords={
__global__,__host__,__device__,__constant__,__shared__,
gridDim,blockIdx,blockDim,threadIdx,
char1,char2,char3,char4,
uchar1,uchar2,uchar3,uchar4,
short1,short2,short3,short4,
ushort1,ushort2,ushort3,ushort4,
int1,int2,int3,int4,
uint1,uint2,uint3,uint4,
long1,long2,long3,long4,
ulong1,ulong2,ulong3,ulong4,
longlong1,longlong2,
float1,float2,float3,float4,
double1,double2,
dim1,dim2,dim3,dim4,
tex1Dfetch,tex1D,tex2D,tex3D,
__float_as_int,__int_as_float,
__float2int_rn,__float2int_rz,__float2int_ru,__float2int_rd,
__float2uint_rn,__float2uint_rz,__float2uint_ru,__float2uint_rd,
__int2float_rn,__int2float_rz,__int2float_ru,__int2float_rd,
__uint2float_rn,__uint2float_rz,__uint2float_ru,__uint2float_rd,
__fadd_rz,__fmul_rz,__fdividef,
__mul24,__umul24,__mulhi,__umulhi,__mul64hi,__umul64hi,
min,umin,fminf,fmin,max,umax,fmaxf,fmax,
abs,fabsf,fabs,rsqrtf,rsqrt,sqrtf,sqrt,__sinf,sinf,sin,__cosf,cosf,cos,
__sincosf,sincosf,sincos,__expf,expf,exp,__logf,logf,log,
__syncthreads,
__threadfence,
__threadfence_block
}
}

\lstdefinelanguage{pseudoCUDA}
[ISO]{C++}
{
morekeywords={
global,host,device,
function,
threadIdx,
threadCount,
blockIdx,
blockCount,
synchronize,
atomic,
in,
do
}
}

\lstdefinelanguage{Impala}
[ISO]{C++}
{
morekeywords={
def, kind, errorType, typeof
}
}

\definecolor{deepcpsstage}{rgb}{0.85,0.45,0.2}
\definecolor{deepcpspseudo}{rgb}{0,0.2,0.85}
\definecolor{langdslarg}{rgb}{0.2,0.55,0.75}
\definecolor{langdsldeepcps}{rgb}{0,0.3,0.1}

\lstdefinelanguage{deepcps}
{
morekeywords={
fix, in, let, stage, always, never,
bool, int, float, fn, meta, function, grammar, epsilon
},
moredelim=[is][\color{deepcpsstage}\ttfamily]{'}{'},
moredelim=[is][\color{deepcpspseudo}\ttfamily\slshape]{"}{"},
moredelim=[is][\ttfamily\slshape]{↔}{↔},
moredelim=[is][\color{langdslarg}\ttfamily]{|}{|},
moredelim=[is][\color{langdsldeepcps}\ttfamily]{▼}{▼},
literate={▲}{{\$}}1 {""}{{"}}1 {''}{{'}}1 {||}{{|}}1
}

\definecolor{langdslcode}{rgb}{0.10,0.25,0.65}

\lstdefinelanguage{langdsl}
{
frame=single,
morekeywords={
language, fragment, keyword, symbol, token, ignore arguments, rule, action, epsilon
},
moredelim=[is][\color{langdslcode}\ttfamily\slshape]{<}{>},
moredelim=[is][\color{deepcpspseudo}\ttfamily\slshape]{""}{""},
}

\definecolor{deepcpsstage}{rgb}{0.65,0.25,0}

\lstdefinelanguage{cstage}[ISO]{C++}
{
morekeywords={fn,stage,def,var},
moredelim=[is][\color{deepcpsstage}\ttfamily]{'}{'}
}

\lstdefinelanguage{MetaML}[]{ML}
{
moredelim=[is][\color{deepcpsstage}]{|}{|},
moredelim=[is][\bfseries\color{deepcpsstage}]{!}{!},
morekeywords={lift,run}
}

\lstdefinelanguage{P}
[]{Impala}
{
morekeywords={
thread, task, shared, solid
},
deletekeywords={global}
}

\lstdefinelanguage{PRT}
[]{P}
{
morekeywords={
persistent, closure, functor
}
}

\newcommand{\inline}[1]{\lstinline[basicstyle=\ttfamily\footnotesize]~#1~}
\newcommand{\tinline}[1]{\text{\lstinline[basicstyle=\ttfamily\footnotesize]~#1~}}
\newcommand{\ignore}[1]{}

\newcommand{\fun}{\lambda\overline{x} . b}
\newcommand{\sfun}{\lambda\Stage{y}\overline{x} . b}
\newcommand{\AND}{\ \texttt{and}\ }
\newcommand{\OR}{\ \texttt{or}\ }
\newcommand{\NOT}{\texttt{not}\ }

\newcommand{\Stage}[1]{\texttt{[}#1\texttt{]}}

\newenvironment{itemize*}
{ \begin{list}{\labelitemii}{
    \setlength{\topsep}{0pt}
    \setlength{\parskip}{0pt}
   \setlength{\itemsep}{-3pt}
    \setlength{\leftmargin}{1em}
}}
{\end{list} }

\newenvironment{enumerate*}
{ \begin{list}{\arabic{enumi}.}{\usecounter{enumi}
    \setlength{\topsep}{0pt}
    \setlength{\parskip}{0pt}
   \setlength{\itemsep}{-1pt}
    \setlength{\leftmargin}{1.4em}
}}
{\end{list} }

\DeclareFloatingEnvironment[
    fileext=lop,
    name=Listing]{code}
\DeclareCaptionSubType{code}

\newcommand{\highlightcode}[6]{
\shade [left color=#1!30, middle color=white, shading=axis,shading angle=45] (#2,#3) rectangle (#4,#5);
\draw[color=#1](#2,#3) -- (#4,#3);
\draw[color=#1](#2,#3) -- (#2,#5);
\node[anchor=north east,inner sep=0.3] at (#4,#3) {\color{#1} \emph{#6}};
}

\newacro{CPS}{Continuation Passing Style}
\newacro{LMS}{Lightweight Modular Staging}
\newacro{DSL}{Domain-Specific Language}

\newcommand{\includeComments}{%
\newcommand{\createComment}[3]{%
\expandafter \newcommand \csname
##1\endcsname[1]{{{\color{##3}\textsf{\textbf{[##2####1]}}}}}%
}}
\newcommand{\includeRemarks}{%
\newcommand{\createRemark}[3]{%
\expandafter \newcommand \csname
##1\endcsname{{{\color{##3}\textsf{\textbf{[##2]}}}}\xspace}%
}}
\newcommand{\noComments}{%
\newcommand{\createComment}[3]{%
\expandafter \newcommand \csname ##1\endcsname[1]{}%
}}

\newcommand{\noRemarks}{%
\newcommand{\createRemark}[3]{%

\expandafter \newcommand \csname ##1\endcsname[1]{}%
}}

\includeComments
\includeRemarks
\createComment{todo}{}{red}
\createRemark{tocite}{CITE}{red}

\hypersetup{linkcolor=black,citecolor=black,urlcolor=black,colorlinks=true}

\renewcommand{\P}[1]{\texttt{\scriptsize P\ref{lst:p#1}}}

\renewcommand{\topfraction}{.9}

\renewcommand{\floatsep}{8pt}
\renewcommand{\textfloatsep}{10pt}
\tikzfading[name=fade right,
  left color=transparent!0, right color=transparent!100] 

\bibliography{literature}
	
\begin{document}

\maketitle
\begin{abstract}

When creating a new domain-specific language (DSL) it is common to embed it as a part of a flexible host language,
rather than creating it entirely from scratch.
The semantics of an embedded DSL (EDSL) is either given directly as a set of functions (shallow embedding),
or an AST is constructed that is later processed (deep embedding).
Typically, the deep embedding is used when the EDSL specifies domain-specific optimizations (DSO) in a form of AST transformations.

In this paper we show that deep embedding is not necessary to specify most optimizations.
We define language semantics as \emph{action functions} that are executed during parsing.
These actions build incrementally a new, arbitrary complex program function.

The EDSL designer is able to specify many aspects of the semantics as a runnable code,
such as variable scoping rules, custom type checking, arbitrary control flow structures, 
or DSO.
A sufficiently powerful staging mechanism helps assembling the code from different actions,
as well as evaluate the semantics in arbitrarily many stages.
In the end, we obtain code that is as efficient as one written by hand.

We never create any object representation of the code.
No external traversing algorithm is used to process the code.
All program fragments are functions with their entire semantics embedded within the function bodies.
This approach allows reusing the code between EDSL and the host language,
as well as combining actions of many different EDSLs.

\end{abstract}


%




\section{Introduction}
\label{sec:introduction}

%

Creating a new DSL from scratch is not an easy task.
Even with the help of parser generators, such as YACC~\cite{yacc},
the developer still needs to handle several compiler-related tasks such as creating and transforming an abstract syntax tree
or implementing variable lookup rules or a type system.

An alternative approach is to \emph{embed a DSL} -- define it as part of a flexible, general-purpose \emph{host language}.
The DSL is defined by a set of functions and other constructs giving the user a perception of programming in a new language.

In this paper we focus on \emph{action code} -- pieces of code that define the semantics of the language.
The simplest approach for the action is to contain the intended domain program behavior directly. 
When the associated syntactic construct is encountered by the interpreter or parser, the domain program is executed immediately.

Alternatively, one uses some form of \emph{staging} -- a user-guided partial evaluation mechanism where the programmer annotates the code, guiding the interpreter to execute the code out of the normal order.
In the context of a DSL definition, staging is used to defer the execution of the domain program. 
Most typical approaches of staging, such as Lightweight Modular Staging (LMS)~\cite{LMS}, produce a program as an Abstract Syntax Tree (AST).
The generated program, represented as an object, can be then traversed, analyzed, optimized and later compiled and run separately.


One embeds a DSL to simplify the definition creation, and to reuse code between the host language and the DSL.
However, when the code is represented as an AST these traits are lost.
The AST traversal and transformations must once again be defined.
Moreover, the host language can no longer be seamlessly integrated with the produced program:
The host code cannot be directly reused in the domain and the implementation details of the embedding can become visible to the end user~\cite{YinYang}.

\subsection*{Contributions}

In this work we show a novel way to define language semantics without representing it as an AST object and traversing it separately.
Despite that constraint, the language should still be able to produce highly efficient code.

We create \emph{fragment functions} which conceptually act as tree nodes.
Unlike existing solutions, however, their behavior is defined entirely by their body, without any external traversing algorithm.
A single node can contain arbitrary code and realize a complex language logic.
We show:
\begin{itemize}
\item How to define and connect these fragment functions.
\item How to use staging to remove overhead and specify early-stage optimizations.
\item How to define tasks which are typically realized through AST traversing algorithms.
This includes basic operations, such as variable name lookup rules, as well as more advanced domain-specific optimizations.
\end{itemize}

Our solution has simplicity of shallowly embedded DSL and the power of deep embedding.

\section{Related Work}
\label{sec:related}

\begin{figure*}[tb]
\begin{tikzpicture}[xscale=0.1685,yscale=-0.425]
\tikzset{verge/.style={-,decorate, decoration={snake}, line width=0.3mm, dash pattern={on 0.5mm off 0.6mm}}}
\node[inner sep=0,anchor=north west] at (0,0) {
\begin{tabularx}{\textwidth}{|r|c|X|l|}\cline{1-4}
\textbf{Embedding type} & \textbf{Staging method} & \textbf{Host language/system} & \textbf{Example languages created with it}\\\cline{1-4}
standalone     &\multirow{2}{*}{structural staging} & LLVM~\cite{LLVM} & clang~\cite{clang-llvm}, Kaleidoscope\footmark{foot:kaleidoscope}, Stacker\footmark{foot:stacker}\\\cline{1-1}\cline{3-4}
deep embedding\hspace{3mm} & & Lisp~\cite{LispQuasi}, Racket~\cite{Racket} & Pollen\footmark{foot:pollen} \\\cline{2-4}
 &\multirow{2}{*}{functional staging}& MetaML~\cite{MetaMLFirst}, MetaOCaml& \cite{MetaML-DSL}, QBF DSL~\cite{MetaOCaml-Haskell-C-DSL} \\\cline{3-4}
 & & LMS~\cite{LMS} & \cite{LMSforDSL} \\\cline{2-4}
 & dynamic staging & DeepCPS~\cite{DeepCPS} & this paper \\\cline{2-4}
 & automated staging & Impala~\cite{Impala} & AnyDSL\footmark{foot:anydsl} \\\cline{2-4}
shallow embedding\hspace{3mm} & no staging & Haskell & ImplicitCAD\footmark{foot:implicitcad} \\\cline{1-4}
\end{tabularx}
};
\draw [stealth-,line width=0.5mm] (18.4,2) -- (18.4,5);
\draw [-stealth,line width=0.5mm] (18.4,6) -- (18.4,8);
\draw[verge] (0,5.5)--(20.5,5.5);
\end{tikzpicture}
\vspace{-5mm}
\caption{\label{fig:stage_to_embed}
The relation between different staging methods and language creation types.
}
\vspace{-3mm}
\end{figure*}

\foottext{foot:kaleidoscope}{Kaleidoscope: Code generation to LLVM IR,\\ http://llvm.org/docs/tutorial/LangImpl3.html, retrieved on 12.05.2015}
\foottext{foot:stacker}{Stacker: An Example Of Using LLVM,\\ http://llvm.org/releases/1.1/docs/Stacker.html, retrieved on 2.12.2016}
\foottext{foot:pollen}{Pollen: the book is a program,\\ http://docs.racket-lang.org/pollen, retrieved on 2.12.2016}
\foottext{foot:anydsl}{AnyDSL -- A Framework for Rapid Development of Domain-Specific Libraries, https://anydsl.github.io/, retrieved on 2.12.2016}
\foottext{foot:implicitcad}{ImplicitCAD, http://www.implicitcad.org/, retrieved on 09.06.2016}

\subsection{Staging}

We call \emph{staging} any method that allows symbolic computation ``under a lambda'', that is:
performing $\beta$-reduction within bodies of functions that have not been invoked yet.
We define the following types of staging:

The simplest, most crude approach is \emph{textual staging}, where a program is composed from strings containing pieces of source code.
Textual staging provides no correctness guarantees about the generated program and has limited mechanisms for communication between the stages.

In \emph{structural staging}, the program code is represented explicitly as a data structure, typically as an AST or graph.
The structure can be created explicitly, for example through the LLVM instruction builders.
Alternatively, the process can be hidden behind overloaded functions, operators, or templates~\cite{CExpr}.
Data exchange between stages is inconvenient since it requires conversion from values to structure nodes and vice-versa.
Moreover, nested staging requires a structure to describe a program building another structure,
making such approach inconvenient to use.

\emph{Functional staging} is a case of structural staging where the structure is represented by ordinary functions~\cite{Reynolds75, Tagless}.
Such functions can be combined to form bigger pieces of code.
The piece of code is executed by simply invoking the function representing it.
In Lightweight Modular Staging (LMS)~\cite{LMS} these builder functions are hidden by overloading ordinary functions over a higher-kinded type \inline{Rep[T]}.
Consequently, the staged code does not differ much from the normal code.
Unlike earlier approaches, communication is facilitated through an implicit promotion of values from \inline{T} to \inline{Rep[T]} where needed.
Still, the structural staging is not entirely hidden and sometimes it leaks to the programmer.
As a result, they must be aware of the internal usage of staging within the builders~\cite{YinYang}.

Languages such as `C~\cite{TickC} or MetaML~\cite{MetaML} use a dedicated syntax to represent code objects.
In MetaML, pieces of code can be combined, spliced, and executed.
Values can be used in many stages through splice and lift operators.
MetaML allows for arbitrarily deep nesting of staging.
An enhanced type system is able to guarantee correctness of staging~\cite{MetaMLIdealized}.

All the above techniques represent the staged code as objects within the host language.
With \emph{dynamic staging}, realized by the DeepCPS language, this view is completely abandoned~\cite{DeepCPS}.
There is no inherent difference between the code in different stages.
Staging is controlled by first-class \emph{staging parameters} that control the execution order by annotating instructions.
Symbolic computation and deferred execution is implicitly achieved when execution occurs under a lambda.

Impala~\cite{Impala} has a similar view on staging.
The language is less verbose and more automatic in selecting what parts of code are executed based on a single user annotation.
Unfortunately, it is not easy to parametrize the staging and have it dependent on arbitrary computation.

We build our work on top of DeepCPS, making use of the language staging flexibility.
We benefit from the lack of any inherent distinction between stages,
as it removes the difference between the immediate action code and the program that is being built.

\begin{figure*}[tb]
\begin{tabularx}{\textwidth}{lp{4.4cm}Xr}
Basic syntax: & $\sfun$ & \inline{(x$_1$, x$_2$, ...)'[y]' \{ $\mathit{b}$\}} & (lambda function) \\
& $\top$ $\bot$ $\AND$ $\OR$ $\NOT$ & \inline{'always'} \inline{'never'} \texttt{\& | !} & (staging constants, operators) \\
& $\Stage{e}v\ \overline{v}$ &
    \inline{'@e:' v v$_1\ $ v$_2\ $ ...} &
    ($\lambda$ body: application) \\
& $\Stage{e}\texttt{fix}\ [y]x = v\ \texttt{in}\ b$ &
    \inline{'@e:' fix '[y]' x} $v$ $b$&
    \text{($\lambda$ body: fix-point combinator)} \\
Syntactic sugar:\hspace{-3mm} & $\lambda\Stage{y}\overline{x} . \Stage{y} v \overline{v}$ &
    \inline{(x$_1$, x$_2$, ...)} \inline{$\ $\{ v v$_1\ $ v$_2\ $ ... \}} &
    (natural staging) \\
& $\Stage{e}\ \left(\lambda\Stage{y}x . b\right)\ v$ &
    \inline{'@e:' let '[y]' $\ x$ $\ v$ $\ b$} &
    (let construct) \\
& $\Stage{e}v\ \overline{v} \left(\lambda[y]x.b\right)$ & \inline{'@e:' v v$_1\ $ v$_2\ $ ... (x$_1$, x$_2$, ...)'[y]' $\mathit{b}$} & (last argument) \\
& $\left(\lambda [y] x . b\right)expr \stackrel{\beta}{\longrightarrow} \left(\lambda [y] x . b\right)v$ & \inline{"expr" (x)'[y]' $\ b$} & (non-CPS expression $expr$) \\
\end{tabularx}
\vspace{-1mm}
\caption{Comparison between lambda calculus extended with dynamic staging (left), and the actual syntax of DeepCPS (right).\label{fig:deepcps_syntax}}
\vspace{-3mm}
\end{figure*}

\subsection{Language Embedding}
Languages can be created as standalone products, or as a library in a flexible host language such as Haskell, ML, or Scala.
The second approach is referred as \emph{language embedding} as it defines the DSL within the \emph{host language}.

Such an approach reduces the complexity of the language creation.
Most of the compiler-specific challenges, such as having a sound type system, are already resolved in the host language and can be reused in the DSL.
The author can focus only on aspects that are truly unique for the domain.


Having the common base language, DSLs can use it to relay information between the domains and the host language.
For this reason, this approach is common when defining small, embedded DSLs (EDSLs).
For example, Haskell has been used extensively to define geometric operations~\cite{Experiment-Haskell},
CAD operations\footmark{foot:implicitcad},
COM component scripting~\cite{EDSL-COM-in-Haskell},
hardware design~\cite{EDSL-Lava}, or
server-side web scripting~\cite{EDSL-web, EDSL-web-WASH}.

%

%
Most EDSLs focus on semantics, while inheriting the syntax of the host.
The semantics is defined by pieces of code, that we refer to as \emph{actions}. 
Depending on its contents we distinguish two forms of embedding:
\begin{itemize}
\item In \emph{shallow embedding} the associated semantics is represented directly in the action and is executed immediately when the syntactic construct is encountered.
\item In \emph{deep embedding} a program structure representing the intended behavior is created, that can later be translated, optimized, and run separately.
\end{itemize}

This distinction is closely related to the form of staging used within the actions.
Typical shallow embedding uses no staging and the code is executed together with the embedded syntax representing it.
Deep embedding on the other hand uses structural staging to generate a program based on the syntax.
Direct structural staging is still very common, as in Haskell/DB~\cite{EDSL-HaskellDB} or Scala~\cite{EDSL-Scala}.
However, more advanced staging approaches become popular in context of EDSLs as well, using LMS or MetaML~\cite{LangVirt, Spiral, MetaML-DSL}. 

This relation between the type of staging and the depth of the embedding, as depicted in \autoref{fig:stage_to_embed}, are of particular interest to us.
In our work we provide a solution which, due to the flexibility of staging, is on the verge between what can be considered shallow and deep embedding.

\section{Dynamic Staging}
\label{sec:dynstage}

Before we present our main work let us quickly summarize the dynamic staging and DeepCPS language as our work heavily utilizes it.
The full information, including the formal definition, can be found in~\cite{DeepCPS}.

\subsection{DeepCPS}

DeepCPS is a functional language operating entirely in the Continuation Passing Style (CPS)~\cite{AppelCPS}.
In \autoref{fig:deepcps_syntax} we summarize the syntax of DeepCPS.
The basic lambda value $\fun$ is represented as \lstinline|(x$_1$,x$_2$,...){$\mathit{b}$}|.
If the lambda is the last argument in an application -- a scenario common in the CPS programming -- the braces surrounding the body $b$ can be omitted.
This way DeepCPS avoids the deep nesting of parenthesis, typically present in other CPS representations.

Dynamic staging is introduced through the \emph{implicit staging parameter} \inline{'[y]'} that is present in every lambda
and a \emph{staging expression} \inline{'@e:'} in each lambda body.
When a lambda is invoked, the implicit staging parameter is substituted by a staging constant $\top$.
Staging expressions use these staging parameters to form boolean expressions.
When \inline{'@e:'} evaluates to $\top$, the annotated body is considered \emph{active} and is scheduled for the execution.
At the each execution step the deepest active body, containing no nested active bodies, is executed.

Staging variables can be used as normal arguments, and normal variables can appear in staging expressions as well.
Non-stage constants are equivalent to $\top$, while variables that are still represented only as symbolic values are $\bot$.

As syntactic sugar, the staging expression can be omitted,
implicitly staging the body on the lambda it is directly contained in.
When that lambda is invoked, the body is executed as if no staging was present.
We refer to this as \emph{natural staging}.

\subsection{Control Flow}
\label{sec:control_flow}

Control flow in DeepCPS is realized through a single built-in function \inline{if} taking 3 arguments: \inline{cond}, \inline{tb}, \inline{fb}.
Depending on a boolean value \inline{cond}, either the true branch (\inline{tb}) or the false branch (\inline{fb}) continuation is invoked.

\begin{code}
\begin{lstlisting}
fix for(↔!args↔, startval, endval, body, endfor) {
  "startval<endval" (cond)
	if cond () {
	  body ↔!args↔ startval endfor (↔!args2↔)
	  "startval+1" (nextval)
	  for ↔!args2↔ nextval endval body endfor
	} ()
	endfor ↔!args↔
} in ...
\end{lstlisting}
\vspace{-2mm}
\caption{\label{code:for}
A \inline{for} loop defined as a higher-order recursive function, using the DeepCPS syntax.
In each step, the user-provided \inline{body} function is called, passing in and out a tuple of arbitrary arguments \inline{!args}.
Upon reaching the end, the \inline{endfor} continuation is invoked.}
\end{code}

Other flow functions can be defined with the help of \inline{if}, for example as shown in \autoref{code:for}.
The \inline{for} function recursively calls itself, incrementing the initial value \inline{startval} by 1 in each call.
In each iteration, the body is invoked with the current value of \inline{startval}.
The recursion is terminated when the \inline{endval} value is reached.

A parameter preceded by \inline{!} accepts any excessive arguments given to a function, and packs them all into a tuple under the given name.
When, for example \inline{for} is invoked with 6 arguments, the first two are put into the tuple \inline{args}.

The symbol \inline{!} in front of an argument unpacks all elements of a tuple and splices them as an argument list to a function call.
For convenience, all arguments that are packed or unpacked through \inline{!}, are highlighted in italics.

\subsection{Staging Chain}
\label{sec:fragment_chaining}

\begin{code}
\begin{lstlisting}
... '@s1:' ... $\textrm{\emph{code 1}}$ ... '[s2]' ...
...
... '@s2:' ... $\textrm{\emph{code 2}}$ ... '[s3]' ...
...
... '@s3:' ... $\textrm{\emph{code 3}}$ ... '[s4]' ...\end{lstlisting}
\vspace{-2mm}
\caption{\label{code:fragment}The staging chain pattern.
When \inline{'s1'} becomes active, $\textrm{\emph{code 1}}$ is executed.
Its final continuation activates \inline{'s2'}, triggering the execution of the next \emph{code} piece.
This way, all the code pieces \emph{code$_n$} are executed in sequence.
The code pieces can occur at arbitrary locations in the program, even in separate functions, as long as the staging values that form the chain are properly passed.}
\end{code}

One of the benefits of DeepCPS that we rely on in this paper is the ability to form staging chains.
A \emph{staging chain} is a sequence of staging variables \inline{'s1'}, \inline{'s2'}, \inline{'s3'}... where $\color{deepcpsstage}s_i$ stages a piece of code that is ending with a continuation defining $\color{deepcpsstage}s_{i+1}$.
The simplest case is shown in \autoref{code:fragment}, but these pieces can originate from completely independent functions.
A single piece may be encapsulated in a function such as:

\begin{lstlisting}
let fragment('s_n', return) {
  ...	'@s_n': ... $\textrm{\emph{code 1}}$ ... '[s_m]' ...
	return 's_m'
}
\end{lstlisting}
\vspace{-1mm}

\section{Building Code}
\label{ch:building_nostage}

Consider a language designer creating a new DSL.
Let us assume that the grammar for such language already exists.
The precise representation of the grammar does not interest us:
the grammar may be defined in a standalone parser, in a Backus-Naur form~\cite{BNF}, or through a set of parser combinator functions,
or as an embedding in another language.

The grammar contains semantic actions in a form of action functions, executed when the given grammar rule is taken.
We assume that each action has means to communicate with other actions by passing arguments.
We only require that the communication is performed in one direction, in the order that the actions are invoked.
This can be the order of parsing, e.g. by using an L-attributed grammar~\cite{LAttribute}.

The question we answer in this section is how to specify these actions, so that by invoking them we generate a new program code.
We use the DeepCPS syntax, but for the moment use no staging.


\subsection{Immediate Execution}

The most straightforward approach is to provide the intended semantic meaning of the parsed code directly in the action.
Suppose for example, one implements a simple stack-based interpreter of binary arithmetic expressions as in \autoref{code:arith_immediate}:
\begin{code}
\begin{lstlisting}
let push(S, v, return) { return [v, S] }
let pop(S, return) {
  "S[0]" (v)
	"S[1]" (tail)
	return v tail
}
\end{lstlisting}
\vspace{-2mm}
\caption{\label{code:stack}
The \inline{push} and \inline{pop} functions for a stack represented as a nested tuple \inline{[v,tail]}.
Each element \inline{v} is paired with a \inline{tail} representing the rest of the stack.
After each operation, an updated stack is passed into the continuation \inline{return}.}
\end{code}
\begin{itemize}
\item We use a stack \inline{S}, represented as a nested tuple \inline{[v,tail]}. For convenience, we define \inline{push} and \inline{pop} in \autoref{code:stack}.
\item Whenever a number is encountered, it is pushed onto \inline{S}                                                                    .
\item When an operator is encountered, the operands are taken from \inline{S} and evaluated, pushing the result back onto \inline{S}.
\end{itemize}
In addition, we include the \inline{start} action, that initializes the stack as an empty tuple, and \inline{end}, that terminates the whole DeepCPS program through a built-in \inline{▲exit}.

\begin{code}
\begin{lstlisting}
let start(return) { return [] }
let number(S, val, return) { push S val return }
let add(S, return) { pop S (arg2, S)
	                   pop S (arg1, S)
	                   "arg1+arg2" (sum)
	                   push S sum return }
let sub ... let mul ... let div ...
let end(S) { pop S ▲exit }
\end{lstlisting}
\vspace{-2mm}
\caption{Semantic actions for an immediate interpreter of arithmetic expressions.
Each action is represented as a DeepCPS function taking a stack argument \inline{S} and returning its updated version through the continuation \inline{return}.
\label{code:arith_immediate}}
\end{code}

Of course, the downside of this approach is that we actually never generate any code.
Instead, the semantic actions compute the result immediately, as in a plain shallowly embedded DSL without any staging.


\subsection{Building Code in CPS}

\begin{code*}
\begin{subfigure}[b]{0.49\textwidth}
\begin{tikzpicture}[xscale=0.16,yscale=-0.32]
\highlightcode{blue}{2}{1}{50}{7}{builder};
\highlightcode{green}{4}{2}{46}{6}{fragment};
\highlightcode{brown}{6}{3}{41}{5}{core};
\node[inner sep=0] at (20.3,3.4) {
\begin{lstlisting}
let build
  (code, return) { return
	  (next, return) { return
		  (↔!args↔)
			  code ↔!args↔ (↔!args2↔) next ↔!args2↔
	  }
  }
\end{lstlisting}
};
\end{tikzpicture}
\end{subfigure}
\begin{tikzpicture}[xscale=0.16,yscale=-0.32]
\highlightcode{blue}{2}{1}{36}{8}{};
\highlightcode{green}{4}{2}{27}{7}{};
\highlightcode{brown}{6}{3}{27}{5}{};
\node[inner sep=0] at (17.5,3.9) {
\begin{lstlisting}
let merge
  (Fprev, Fnext, return) { return
    (next, return) {
	    Fnext next (core)
		  Fprev core (core2)
			return core2
	  }
  }
\end{lstlisting}
};
\end{tikzpicture}
\vspace{-2mm}
\caption{\label{code:builder_single}
The basic builder and merge functions for our functional code building.
The builder creates a fragment function that encapsulates a piece of arbitrary user code, given as an argument to the builder.
The merge function connects two existing fragment functions \inline{Fprev}, \inline{Fnext},
forming a single new fragment containing all user code from  \inline{Fprev} and \inline{Fnext}.}
\end{code*}

Instead of performing the computation immediately, an action should produce a piece of a code that can be called later.
We achieve that by encapsulating it within lambda functions.
We introduce \emph{builder functions} -- a CPS approach for such encapsulation, consisting of 3 layers of functions:
\begin{itemize}
\item The \emph{core} constitutes a piece of code that we want to generate.
In CPS representation, it is a function taking any number of arguments,
performing its user-defined function and then returning through the continuation with a (possibly different) set of arguments.
\item The core is encapsulated by a \emph{fragment function} which specifies the continuation for the core.
\item Finally, a \emph{builder function} acts as a constructor:
When invoked with an arbitrary code as an argument, it creates a fragment function with the core containing said code.
\end{itemize}

A simple builder from \autoref{code:builder_single} calls its continuation \inline{return} with a newly created fragment function.
By invoking the fragment with a specified continuation \inline{next} we create a core.
The core takes arbitrary many parameters, indicated by \inline{!}, and passes them to the user-defined \inline{code}.
Therefore, in order to form and execute a program one needs to combine:

\vspace{0.5mm}
{\centering builder(code) + fragment(next) + core(args)\\}
\vspace{0.5mm}

Fragment functions are convenient to use because they can be combined together through a \emph{merge function} to form a bigger fragment with the same structure.
The \inline{merge} from \autoref{code:builder_single} takes two fragment functions \inline{Fprev} and \inline{Fnext}.
The core of \inline{Fnext} is given as the continuation for the core of \inline{Fprev}.
This way the two bodies become connected, with contents of \inline{Fnext} appearing after \inline{Fprev}:

\begin{tikzpicture}[xscale=0.1685,yscale=-0.315]
\highlightcode{green}{0}{0}{46}{6}{merged fragment};
\highlightcode{brown}{2}{1}{37}{5}{merged code};
\node[inner sep=0] at (14.5,2.9) {
\begin{lstlisting}
(next, return) { return
	(↔!args↔) 
	  FprevCode ↔!args↔ (↔!args2↔)
		FnextCode ↔!args2↔ (↔!args3↔)
		next ↔!args3↔
}
\end{lstlisting}
};
\end{tikzpicture}

The resulting fragment function has the same structure as a fragment function created by the builder,
and can be further merged with other fragments to build bigger programs.

We must remark however, that the reduction described above is performed within the body of the fragment lambda returned by the \inline{merge} function.
This implies staging. Without it, we actually obtain a result that is only $\beta$-equivalent to the one shown above.
Working with such fragments would incur a major overhead.
We fix this problem using staging in \autoref{ch:building_stage}.

\begin{code}
\begin{tikzpicture}[xscale=0.1685,yscale=-0.33]
\highlightcode{blue}{2}{1}{41}{3}{};
\highlightcode{brown}{8}{1.2}{39}{2}{};
\highlightcode{blue}{2}{6}{41}{12}{};
\highlightcode{brown}{8}{6.2}{39}{11}{};
\node[inner sep=0] at (18.5,7.0) {
\begin{lstlisting}
let number(Fprev, val, return) {
  build (S, cont) { push S val cont }
		    (Fnumber)
	merge Fprev Fnumber return
}
let add(Fprev, return) {
  build (S, cont) {
          pop S (arg2, S)
	        pop S (arg1, S)
	        "arg1+arg2" (sum)
	        push S sum cont
	      } (Fadd)
  merge Fprev Fadd return
}
\end{lstlisting}
};
\end{tikzpicture}
\vspace{-2mm}
\caption{Semantic actions for an interpreter of arithmetic expressions, creating a lambda function with the code containing all the parsed operations.
\label{code:arith_simplebuild}}
\end{code}

With the help of the builders we now redefine our arithmetic expression language.
In \autoref{code:arith_simplebuild}, we no longer execute the code immediately.
Instead, each of the actions creates a new fragment, and merges it with the previous one (\inline{Fprev}).

\subsection{Removing the Builder Overhead}
\label{ch:building_stage}

\begin{code*}
\begin{subfigure}[b]{0.53\textwidth}
\begin{tikzpicture}[xscale=0.1685,yscale=-0.33]
\highlightcode{blue}{2}{1}{47}{8}{};
\highlightcode{green}{4}{2}{46}{7}{};
\highlightcode{brown}{6}{3}{45}{6}{};
\node[inner sep=0] at (22.0,3.9) {
\begin{lstlisting}
let build
  (code, return) { return
	  (next, return) { return
		  ('ct', 'pt', ↔!args↔)
			  '@ct:'   code 'pt' ↔!args↔ ('pt' ↔!args2↔)'[ct]'
				'@next:' next 'ct' 'pt' ↔!args2↔
	  }
  }
\end{lstlisting}
};
\end{tikzpicture}
\end{subfigure}
\begin{subfigure}[b]{0.46\textwidth}
\begin{tikzpicture}[xscale=0.1685,yscale=-0.33]
\highlightcode{blue}{2}{1}{42}{8}{};
\highlightcode{green}{4}{2}{38}{7}{};
\highlightcode{brown}{6}{3}{34}{6}{};
\node[inner sep=0] at (18.5,3.9) {
\begin{lstlisting}
let merge
  (Fprev, Fnext, return)'[bt]' { return
    (next, return) {
	    '@bt:'     Fnext next (core)
		           Fprev body (core2)
			'@return:' return core2
	  }
  }
\end{lstlisting}
};
\end{tikzpicture}
\end{subfigure}
\vspace{-2mm}
\caption{\label{code:builder_single_staged}
The simple builders with dynamic staging. The core uses two fragment chaining patterns \inline{'ct'} and \inline{'pt'}.
Moreover, calling \inline{merge} triggers the \inline{'bt'} stage. The fragments \inline{Fnext} and \inline{Fprev} are merged immediately in the body of lambda \inline{(next, return)...}}
\vspace{-3mm}
\end{code*}

The simple builders we presented so far are lazy in their evaluation:
the merging of the fragments does not occur until the actual program is executed, incurring a potential major overhead.
Instead, we would like to produce any code as if it was written directly in DeepCPS.
The whole layer of builders, fragments, and core functions calls should resolve early, leaving only the user-defined code.
Moreover, we would like to be able to express optimization and staging strategies in the semantic actions as well.

In this section we distinguish three execution phases:
\begin{itemize}
\item The \emph{build-time} (\inline{'bt'}) when the builders are being invoked.
\item The \emph{core-time} (\inline{'ct'}) when all code pieces are already merged into a single fragment and are ready to be transformed into a produced program.
\item The \emph{program-time} (\inline{'pt'}) when the function that we created -- referred as a \emph{program} -- is invoked.
\end{itemize}

We avoid the more common terms compile-time and run-time, because this distinction can be misleading in the context of our builders and dynamic staging.
This is because all the execution occurs in the very same language: DeepCPS.
All code, including the programs we construct, can be executed immediately, even during build-time of another function.
Thus, a program-time for one function can be a build-time for another.
Such flexibility becomes lost when thinking in rigid terms of compilation and final run-time execution.

With the dynamic staging it is possible to achieve our goals.
The staged builders in \autoref{code:builder_single_staged} have the same structure as the original ones from \autoref{code:builder_single}.
The difference is the use of staging within the bodies of the build and merge functions.

In \autoref{code:staged_merge_run} we show how the staging allows the merge to remove the overhead of having layered fragment function calls.
Suppose that \inline{merge} is invoked with two fragments \inline{Fnext} and \inline{Fprev} produced by the builder.
As part of the invocation, the implicit staging parameter \inline{'bt'} becomes $\top$ and triggers immediate execution of the code within the body of the merge fragment (\autoref{code:staged_merge_run:step1}).

\begin{code}
\begin{subfigure}[b]{0.48\textwidth}
\begin{tikzpicture}[xscale=0.1685,yscale=-0.33]
\node[inner sep=0] at (21.0,7.4) {
\begin{lstlisting}
build code1 (Fprev)
build code2 (Fnext)
merge Fprev Fnext (F) ...
\end{lstlisting}
};
\end{tikzpicture}
\vspace{-2mm}
\caption{\label{code:staged_merge_run:step0}
Step 0: Initial setup -- we provide two code pieces into the builder and then try to merge the produced fragments.}
\end{subfigure}
\begin{subfigure}[b]{0.48\textwidth}
\begin{tikzpicture}[xscale=0.1685,yscale=-0.33]
\highlightcode{lime}{0}{0}{51}{15}{merge fragment};
\highlightcode{green}{2}{1}{48}{6}{Fnext};
\highlightcode{brown}{4}{2}{44}{5}{};
\highlightcode{green}{2}{7}{48}{12}{Fprev};
\highlightcode{brown}{4}{8}{44}{11}{};
\node[inner sep=0] at (21.5,7.4) {
\begin{lstlisting}
(next, return) {
  '@$\color{deepcpsstage}\top$:' (next, return) { return
	  ('ct', 'pt', ↔!args↔)
		  '@ct:'   code2 'pt' ↔!args↔ ('pt' ↔!args2↔)'[ct]'
			'@next:' next 'ct' 'pt' ↔!args2↔
  }
  next (core)
  (next, return) { return
	  ('ct', 'pt', ↔!args↔)
		  '@ct:'   code1 'pt' ↔!args↔ ('pt' ↔!args2↔)'[ct]'
			'@next:' next 'ct' 'pt' ↔!args2↔
  }
    core
    (core2)	'@return:' return core2
}
\end{lstlisting}
};
\end{tikzpicture}
\vspace{-6mm}%
\caption{\label{code:staged_merge_run:step1}
Step 1: The \inline{merge} function has been invoked with \inline{Fprev} and \inline{Fnext} arguments produced by the builder.
The respective lambda values replace the symbolic names in the universal merge function.}
\end{subfigure}
\begin{subfigure}[b]{0.48\textwidth}
\begin{tikzpicture}[xscale=0.1685,yscale=-0.33]
\highlightcode{lime}{0}{0}{51}{12}{merge fragment};
\highlightcode{green}{2}{1}{48}{6}{Fprev};
\highlightcode{brown}{4}{2}{44}{5}{};
\highlightcode{brown}{4}{6}{48}{10}{core of Fnext};
\node[inner sep=0] at (21.5,5.9) {
\begin{lstlisting}
(next, return) {
  '@$\color{deepcpsstage}\top$:' (next, return) { return
	  ('ct', 'pt', ↔!args↔)
		  '@ct:'   code1 'pt' ↔!args↔ ('pt' ↔!args2↔)'[ct]'
			'@next:' next 'ct' 'pt' ↔!args2↔
  }
    ('ct', 'pt', ↔!args↔) {
		  '@ct:'   code2 'pt' ↔!args↔ ('pt' ↔!args2↔)'[ct]'
			'@next:' next 'ct' 'pt' ↔!args2↔
		}
    (core2)	'@return:' return core2
}
\end{lstlisting}
};
\end{tikzpicture}
\vspace{-6mm}%
\caption{\label{code:staged_merge_run:step2}
Step 2: The \inline{Fnext} has been invoked and it has immediately returned through the \inline{(core)...} continuation.
The argument \inline{core} is now the core of \inline{Fnext}.}
\end{subfigure}
\begin{subfigure}[b]{0.48\textwidth}
\begin{tikzpicture}[xscale=0.1685,yscale=-0.33]
\highlightcode{lime}{0}{0}{51}{8}{merge fragment};
\highlightcode{brown}{4}{2}{51}{7}{core2, core of Fprev};
\highlightcode{brown}{6}{4}{51}{7}{next, core of Fnext};
\node[inner sep=0] at (21.5,3.9) {
\begin{lstlisting}
(next, return) {
  '@return:' return
	  ('ct', 'pt', ↔!args↔)
		  '@ct:'   code1 'pt' ↔!args↔ ('pt' ↔!args2↔)'[ct]'
			
			'@ct:'   code2 'pt' ↔!args↔ ('pt' ↔!args2↔)'[ct]'
			'@next:' next 'ct' 'pt' ↔!args2↔
}
\end{lstlisting}
};
\end{tikzpicture}
\vspace{-6mm}%
\caption{\label{code:staged_merge_run:step3}
Step 3: \inline{Fprev} has been called with the core of \inline{Fnext} as the \inline{next} argument.
This has caused the \inline{Fnext}'s core to be immediately invoked due to \inline{'@next:'} staging.}
\end{subfigure}
\caption{\label{code:staged_merge_run}
Example run of staged \inline{merge} invoked with fragment function arguments produced by the staged \inline{build}.
}
\end{code}

The \inline{Fnext} is given a symbolic argument \inline{next}.
The function merely invokes the returning lambda continuation, with \lstinline|('ct', 'pt', ↔!args↔){...}| as the \inline{body} argument (\autoref{code:staged_merge_run:step2})

This is followed by the invocation of \inline{Fprev}.
This time however, the fragment argument \inline{next} is a concrete lambda -- the previously returned core of \inline{Fnext}.
As a result, within the body of the fragment \inline{Fprev}, the \inline{next} function is immediately invoked.
Consequently, the whole line:\\
\lstinline|'@next:' next 'ct pt' ↔!args2↔|\\
is replaced with the core of \inline{Fnext}.
Only when that happens, the fragment \inline{Fprev} returns (\autoref{code:staged_merge_run:step3}).

Suppose that we have constructed and merged all fragments representing our program.
What we get is a single fragment function of the form:
\begin{lstlisting}
let Fcomplete(next, return) { return
  ('ct', 'pt', !args)
	  '@ct:'   code1 'pt' ↔!args↔ ('pt2' ↔!args2↔)'[ct2]'
		'@ct2:'  code2 'pt2' ↔!args2↔ ('pt3' ↔!args3↔)'[ct3]'
		'@ct3:'  code3 'pt3' ↔!args3↔ ('pt4' ↔!args4↔)'[ct4]'
		...
		'@next:' next 'ctn' 'ptn' !argsn
}
\end{lstlisting}

We can identify the use of the fragment chaining pattern from \autoref{sec:fragment_chaining} twice, associated with \inline{'ct'} and \inline{'pt'}.

The core-time staging chain \inline{'ct'} is used to stage the calls to the user functions \inline{code$_n$}, effectively executing the contents of the core functions (hence the staging name).
This resolves the process of packing and unpacking parameters into the \inline{!args} tuples.
After the core-time chain is resolved, only the code actually provided by the user remains.

If the user code has no staging (that is -- has only natural staging),
\emph{all} of it is evaluated at core-time.
However, by using the program-time \inline{'pt'} staging chain, execution of parts of the code are delayed and spliced into \inline{Fcomplete}.
The code staged upon \inline{'pt'} constitutes the program we generate.

Notice that the builders themselves never stage upon \inline{'pt'} -- it is merely passed around together with \inline{!args}.
In fact, \inline{'pt'} can be part of \inline{!args}, but we included it explicitly for clarity.
Moreover, \inline{!args} can include more user-defined staging variables.
With these, users are able to define arbitrary additional dynamic staging within their code.

\subsection{Branching via Universal Builders}

The basic builders we work with so far assume that the code requires a single continuation.
We say that these fragment functions are of arity 1.
However, when more continuations are needed, the signature of the fragment changes and the builders have to be adjusted.
For example, when encapsulating a condition function \inline{if} that takes two continuations:
\begin{lstlisting}
build2 (trueCont falseCont) {
	       if cond trueCont falseCont
	     }
	     (F) ...
\end{lstlisting}
one needs to use a version of the builder which creates a fragment function of arity 2, for example defined as \inline{build2}:

\noindent
\begin{tikzpicture}[xscale=0.1685,yscale=-0.33]
\node[inner sep=0] at (22.0,3.9) {
\begin{lstlisting}
let build2
  (code, return) { return
	  (next1, next2, return) { return
		  (↔!args↔)
			  code ↔!args↔ (↔!args2↔) { next1 ↔!args2↔ }
				           (↔!args3↔) { next2 ↔!args3↔ }
	} }
\end{lstlisting}
};
\end{tikzpicture}

Fragment functions of different arity present a problem as they cannot be directly used with the \inline{merge} function presented earlier.
On the other hand, creating a merge function to match fragments of different arity would be ineffective --
the user would have to use several versions of the function, where the distinction could be automated.

\begin{code}
\begin{subfigure}[b]{0.53\textwidth}
\begin{tikzpicture}[xscale=0.1685,yscale=-0.333]
\highlightcode{blue}{2}{1}{50}{20}{builder};
\highlightcode{green}{4}{2}{49}{19}{fragment};
\highlightcode{brown}{8}{15}{48}{17}{core};
\highlightcode{brown}{12}{7}{48}{10}{continuation $\lambda_i$};
\node[inner sep=0] at (25.0,9.9) {
\begin{lstlisting}
let build
  (arity, code, return)'[bt]' { return
    (Snext, return) {
		  '@bt:' for ↔Snext []↔ 0 arity
			  (↔Snext, Tlambda↔, i, break, continue)'[bt]' {
				  '@Snext:' pop Snext (next, Snext)
				  '@bt:' let lambda_i 
					  ('pt', ↔!args2↔)'[ct]'{
						  '@next:' next 'ct' 'pt' ↔!args2↔
						}
				  "concat(Tlambda, [lambda_i])" (Tlambda)
				  continue ↔Snext Tlambda↔
			  }
			(↔Snext, Tlambda↔) //endfor
			push Snext 
			  ('ct', 'pt' ↔!args↔) {
				  '@ct:' code 'pt' ↔!args !Tlambda↔
				} (S)
			'@return:' return S
  } }
\end{lstlisting}
};
\end{tikzpicture}
\end{subfigure}
\begin{subfigure}[b]{0.45\textwidth}
\begin{tikzpicture}[xscale=0.1685,yscale=-0.33]
\highlightcode{blue}{2}{1}{45}{8}{};
\highlightcode{green}{4}{2}{41}{7}{};
\highlightcode{brown}{6}{3}{37}{5}{};
\node[inner sep=0] at (18.5,3.9) {
\begin{lstlisting}
let merge
  (Fprev, Fnext, return)'[bt]' { return 
    (Snext1, return) {
		  '@bt:'     Fnext Snext1 (Snext2)
  		         Fprev Snext2 (Snext3)
			'@return:' return Snext3
    }
  }
\end{lstlisting}
};
\end{tikzpicture}
\end{subfigure}
\vspace{-3mm}
\caption{\label{code:builder_universal}
The universal builders handling fragments of arbitrary arity. The fragments exchange a stack of continuations \inline{Snext}, rather than a single continuation \inline{next}. In the \inline{for} loop, the fragment takes as many elements from the stack as needed, defined by the \inline{arity} parameter, and pass them into the user-defined \inline{code} as continuations.}
\end{code}

Instead of using specialized builder and merge function, in \autoref{code:builder_universal} we introduce a universal solution for any arity.

The main difference is that we no longer pass a single \inline{next} continuation into the fragment functions.
Instead, we pass a stack \inline{Snext}, containing arbitrary many continuations.
We use the same stack implementation as previously in \autoref{code:stack}.

When creating a new fragment through the builder, the user specifies the intended \inline{arity}.
This specifies the number of continuations passed by the builder to the user-provided \inline{code} that follow the standard \inline{!args}.
These continuations are kept in the \inline{Tlambda} tuple, created at build-time in the \inline{for} loop at the beginning of the builder's fragment function.
Each of the \inline{next} values used by these continuations is obtained from the \inline{Snext} stack as soon as it is available.

\begin{code}
\begin{tikzpicture}[xscale=0.1685,yscale=-0.33]
\highlightcode{green}{0}{0}{50}{10}{};
\highlightcode{brown}{3}{4}{49}{8}{core (top stack element)};
\highlightcode{brown}{5}{6}{50}{7}{$\lambda_0$};
\highlightcode{brown}{5}{7}{50}{8}{$\lambda_1$};
\node[inner sep=0] at (23.5,4.9) {
\begin{lstlisting}
(Snext, return) {
  '@Snext:' pop Snext (next, Snext2)
	'@Snext2:' pop Snext (next2, Snext3)
	'@return:' return
	[('ct', 'pt', !args) {
	  '@ct:' code 'pt' !args ![
		 ('pt', !args) { '@next:' next 'ct' 'pt' !args },
		 ('pt', !args) { '@next2:' next2 'ct' 'pt' !args }]
	}, Snext3]
}
\end{lstlisting}
};
\end{tikzpicture}
\vspace{-7mm}
\caption{\label{code:build_arity_2_staged}
A fragment function obtained by invoking the universal \inline{build} with arity 2.}
\end{code}

The universal merge function does not differ much from its simple predecessor.
However, this time we do not pass a single core argument between the fragments, but the whole continuation stack \inline{Snext}.
The code within the universal builder, through the use of \inline{lambda_i} performs the actual merging.
The \inline{merge} function merely specifies which two fragment functions are connected.

Note that the \inline{push} operation on the stack is performed in build-time even if \inline{Snext} is a symbolic value.
This allows us to create stack with top element known, and unknown remaining tail.
This can be seen for example in \autoref{code:build_arity_2_staged}.
Such element can be later used in another fragment to resolve the merge, even if the whole stack is not provided.

The logic of the construction through these builders is from back to front: when the continuations representing the ending part of the program are known, they are put on the stack and can be merged with a new fragment of code preceding it.
Despite that it is not necessary to call these functions in that one particular order.
Thanks to the operations on incomplete stack, the \inline{Fnext} can be successfully merged to \inline{Fprev} even if the former is just a short piece of code without any known continuations.

\section{Examples}
\label{ch:usage}

In the previous section we have shown how arbitrary code can be created using a functional approach.
We have explained how dynamic staging removes the overhead of the layered construction.
Let us now show how the builders can be used in practice.

\subsection{Arithmetic Expressions}
\label{ch:usage:arith}

At the beginning of \autoref{ch:building_nostage}, in \autoref{code:arith_immediate} we provide semantic actions for parsing arithmetic expressions.
In there, however, all the computations are performed immediately during parsing. With the help of the builders, the complete set of semantic actions would look like in \autoref{code:arith_built}.

\begin{code}
\begin{tikzpicture}[xscale=0.1685,yscale=-0.33]
\highlightcode{blue}{2}{1}{50}{4}{};
\highlightcode{brown}{4}{2}{44}{3}{};

\highlightcode{blue}{2}{6}{40}{10}{};
\highlightcode{brown}{10}{6.2}{38}{9}{};

\highlightcode{blue}{2}{13}{40}{20}{};
\highlightcode{brown}{10}{13.2}{38}{19}{};

\highlightcode{blue}{2}{24}{40}{28}{};
\highlightcode{brown}{10}{24.2}{38}{27}{};

\node[inner sep=0] at (20.0,16.95) {
\begin{lstlisting}
let start(return) { 
  build 1
		('pt', ↔exit,↔ cont) { cont 'pt' ↔exit []↔ }
	  return
}
let number(Fprev, val, return) {
  build 1 ('pt', ↔exit, S,↔ cont) {
		        push S val (S)
			      cont 'pt' ↔exit S↔
		      } (F)
	merge Fprev F return
	}
let add(Fprev, return) {
  build 1 ('pt', ↔exit, S,↔ cont) {
                 pop S (arg2, S)
	               pop S (arg1, S)'[ct]'
	          '@pt:' "arg1+arg2" (sum)'[pt]'
	          '@ct:' push S sum (S)
			           cont 'pt' ↔exit S↔
          } (F)
	merge Fpref F return
}
let sub... let mul... let div...
let end(Fprev, return) {
  build 0 ('pt', ↔exit, S↔) {
		            pop S (v, S)
			      '@pt:' exit v
		      } (F)
	merge Fpref F (Fcomplete)
	Fcomplete [] (program)
	return (exit)'[pt]' {
		       '@program:' program 'pt' exit
		     }
}
\end{lstlisting}
};
\end{tikzpicture}
\vspace{-7mm}
\caption{Semantic actions for an interpreter of arithmetic expressions.
Each action builds a fragment with the provided user code. The user code employs a number stack to accumulate intermediate values.
The ending action ``seals'' the chain with a fragment of arity 0, taking no more constructed continuations.
\label{code:arith_built}}
\end{code}

The \inline{start} action is executed first, at the beginning.
Apart from the continuation, it takes the function-time staging variable, 
and the \inline{exit} continuation that should end the program we are building.
Moreover, we initialize the number stack with an empty value \inline{[]}.

The \inline{number} and \inline{add} actions construct a fragment of arity 1, similarly to what was done in \autoref{code:arith_simplebuild}.
All operations on the number stack are performed at core-time, even if the numeric values, such as \inline{sum} are not known at that time.
Only the actual addition is delayed to be part of the program we generate, executing at program-time.

Finally, in the \inline{end} semantic action we seal the fragments.
That means, we provide the last fragment with arity 0, which calls \inline{exit} to end the program.
The fragment representing the complete program is invoked with an empty stack as \inline{Snext},
returning a \inline{program} which is a naked user code without any encapsulation.
By invoking the program under the lambda \inline{(exit)'[pt]'...} we trigger the core-time stage,
leaving only the code staged upon \inline{'pt'} in the body of  that lambda.

\begin{code}
\begin{tikzpicture}[xscale=0.1685,yscale=-0.33]
\highlightcode{blue}{2}{1}{45}{5}{};
\highlightcode{brown}{10}{1.2}{42}{4}{};

\highlightcode{blue}{2}{7}{45}{9}{};
\highlightcode{brown}{10}{7.2}{42}{8}{};

\highlightcode{blue}{2}{12}{45}{18}{};
\highlightcode{brown}{10}{12.2}{42}{17}{};

\node[inner sep=0] at (20.0,13.5) {
\begin{lstlisting}
let start(return) { 
  build 1 ('pt', ↔exit,↔ cont) {
		        cont 'pt' ('pt', arg)
			      exit arg
		      } return
}
let number(Fparent, val, return) {
  build 0 ('pt', ↔parent↔) { parent 'pt' val }
		(F)
	merge Fparent F return
	}
let add(Fparent, return) {
  build 2 ('pt', ↔parent↔, left, right) {
		             left 'pt' ('pt', arg1)
			           right 'pt' ('pt', arg2)
	          '@pt:' "arg1+arg2" (sum)'[ft]'
	          '@ct:' parent 'pt' ↔sum↔
          } (F)
	merge Fparent F return
}
let sub... let mul... let div...
let end(Fparent, return) { 
	Fparent [] (program)
	return (exit)'[pt]' {
		       '@program:' program 'pt' exit
		     }
}
\end{lstlisting}
};
\end{tikzpicture}
\vspace{-3mm}
\caption{Semantic actions for the same interpreter of arithmetic expressions.
Here, however, the fragments are connected in an AST-like fashion.
Operators are inner nodes of arity 2, while numbers are leafs with arity 0.
\label{code:arith_AST}}
\end{code}

The presented approach using a number stack is not the only way to realize an arithmetic expression semantics.
Alternatively, the fragments can be connected in an AST-like fashion as in \autoref{code:arith_AST}.
The action associated with the operator creates a fragment function of arity 2.
Each child fragment represents a subexpression on the left and right side of the operator, respectfully.
A literal number, on the other hand, acts as a leaf in the tree of fragments and has arity 0.

In both approaches, we rely on the language grammar to invoke the semantic actions in the correct order.
In particular, the grammar handles the precedence of the operators.

Regardless of whether we use a number stack, or AST-like fragments, we obtain the same end result.
This is because the merging and calls into each user code function is resolved early. 
Only instructions in the \inline{'pt'} chain remain. They are the same in both approaches.
If, for example, we try to parse \inline{1+4*2+3}, the produced program for both cases is exactly:

\begin{lstlisting}
(exit)'[pt]' {
  '@pt:' "4*2" (prod)'[pt]'
  '@pt:' "1+prod" (sum)'[pt]'
  '@pt:' "sum+3" (sum)'[pt]'
	'@pt:' exit sum
}
\end{lstlisting}

\subsection{Custom Environment}
\label{ch:usage:env}

Suppose that a DSL requires name binding.
With functional code building we do not have to rely on a single approach or follow the rules of the host language.
Name binding can be handled explicitly by the user code, possibly in an early stage to avoid program-time overhead.

In order to manage the names, we need a map $M$ that maps string names to values.
We do not care how $M$ is implemented. 
It may be implemented in a purely functional manner, with each operation producing a new $M$ value.
It may use stateful memory operations.
It may even use external C functions for efficiency purposes.
It may also feature a mechanism to reference other maps to support nested scopes.
Since dynamic staging is explicit and does not rely on program analysis, any implementation can be used with our approach.

We only require that
\inline{"M.insert(name, value)"} should bind, possibly symbolic, \inline{value} to \inline{name} in the map $M$,
while \inline{"M.lookup(name)"} should return the previously stored \inline{value}.

\begin{code}
\begin{tikzpicture}[xscale=0.1685,yscale=-0.33]
\highlightcode{blue}{2}{1}{40}{5}{};
\highlightcode{brown}{10}{1.2}{37}{4}{};

\highlightcode{blue}{2}{8}{40}{13}{};
\highlightcode{brown}{10}{8.2}{37}{12}{};

\node[inner sep=0] at (19.0,7.5) {
\begin{lstlisting}
let reference(Fparent, name, return) {
  build 0 ('pt', ↔M, parent↔) {
		        "M.lookup(name)" (value)
		        parent 'pt' valul
	        } (F)
	merge Fparent F return
}
let bind(Fparent, name, return) {
  build 1 ('pt', ↔M, parent↔, child) {
		        child 'pt' ('pt', arg)
			      "M.bind(name, arg)" ()
	          parent 'pt' ↔arg↔
          } (F)
	merge Fparent F return
}
\end{lstlisting}
};
\end{tikzpicture}

\vspace{-2mm}
\caption{Extending the AST-based interpreter of arithmetic expressions to support name binding.
The parameter \inline{M}, acting as a binding map, is passed through all user code functions.
The insertion and lookup is performed at build-time, leaving no trace of it at function-time.
\label{code:arith_env}}
\end{code}

As an example, we can now augment our arithmetic expression language to support name binding, as in \autoref{code:arith_env}.
All map operations are performed at core-time, allowing the map itself to be removed before the program-time.

This is the simplest example of an \emph{auxiliary computation} that can be added into our functional code builders.
Such computation does not constitute the main program logic.
Instead, it is an additional support code that can be used for management, verification or domain-specific optimization.

\subsection{Domain-Specific Optimization}
\label{ch:usage:matrix}

Typically, an optimization is specified as a separate pass over the program structure.
In such a pass, characteristic patterns are found and the AST is transformed, often violating the inherent node semantics.

With the functional builders and dynamic staging the optimization can often be specified as a function as well.
Consider for example our arithmetic expression language operating on matrices of arbitrary dimensions.
Each matrix \inline{M} is represented by its dimensions \inline{"M.dim"} known statically, and its contents \inline{"M.data"} that are known only at program-time.

Consider a problem where a sequence of matrices has to be multiplied.
Finding the optimal sequence of multiplication is a classic problem that can be solved through dynamic programming~\cite[Chap. 15.2]{Cormen}.
Suppose that the algorithm is implemented in a function \inline{"matrix_chain_order(d)"}, which for given tuple $d$ of matrix dimensions computes a sequence describing the optimal multiplication order.
As before, the implementation of such function does not concern us. 

\begin{code}
\begin{tikzpicture}[xscale=0.1685,yscale=-0.33]


\node[inner sep=0] at (19.0,7.5) {
\begin{lstlisting}
let mulStart(Fparent, return) {
  build 1 ('pt', ↔parent↔, children) {
	          children 'pt' [] parent
	        } (F)
	merge Fparent F return
}
let mulNext(Fparent, return) {
  build 2 ('pt', ↔Ms, parent↔, child, continue) {
	           child 'pt' ('pt', M)
						 "concat(Ms, [M])" (Ms)
						 continue 'pt' Ms parent
	        } (F)
	merge Fparent F return
}
let mulEnd(Fparent, return) {
  build 0 ('pt', ↔Ms, parent↔) {
	  "count(Ms)-1" (opCount)
		"get M.dim from all in Ms" (dims)
		"matrix_chain_order(dims)" (order)
		for 'pt' Ms 0 opCount
		  ('pt', Ms, i, break, continue) {
		       "order[i]" (idx)
					 "Ms[idx].data" (M1)
					 "Ms[idx+1].data" (M2)'[ct]'
			'@pt:' "M1*M2" (prod)'[pt]'
			'@ct:' "[Ms[idx+1].dim.x,Ms[idx].dim.y]" (dim)
			     "Ms.delete(idx..idx+1)" (Ms)
			     "Ms.insert(idx,[dim,prod])" (Ms)
			     continue 'pt' Ms
		  } ('pt', Ms)
		"Ms[0]" (finalProd)
		parent 'pt' finalProd
	} (F)
	merge Fparent F return
}
\end{lstlisting}
};
\end{tikzpicture}
\vspace{-2mm}
\caption{\label{code:arith_matrix}
Chain matrix multiplication as an example of domain-specific optimization.
The \inline{mulNext} action collects all the matrices, but the classic algorithm is contained entirely within the \inline{mulEnd} action. 
There, the optimal order is found in \inline{"matrix_chain_order"}.
Based on the result, matrices are multiplied in a \inline{for} loop.
All the optimization is performed in the core-time staging chain.
At program-time the loop is unrolled and only the multiplication operations remain, referencing directly the right matrices.
}
\end{code}

In \autoref{code:arith_matrix} we show how such optimization can be specified in the semantic actions.
The \inline{mulStart} is invoked once at the beginning of the sequence, initializing the list of matrices \inline{Ms} with an empty tuple.
The \inline{mulNext} is used for every member of the matrix chain and accepts 2 continuations.
The \inline{child} represents a matrix subexpression yielding a single result that needs to be multiplied.
The \inline{continue} refers to the remainder of the matrix chain.

Finally, at the end, in \inline{mulEnd} we have all the data to launch the algorithm.
The dimensionality information is extracted from \inline{Ms} and the optimal multiplication order is found.
With each iteration of the \inline{for} loop a single line of code is produced, staged for program-time,
which multiplies the matrices selected by the optimal order.
All other support code: launching the algorithm, managing the matrix tuple and computing the dimensions of the result are executed early, at core-time.

\section{Discussion}
\label{sec:discussion}

Let us discuss how our functional approach for code building relates to both the shallow and the deep language embeddings.

Having the semantic action, containing the intended program logic represented as a function is typical for shallow embedding.
Every host language function can be used directly in the DSL and vice-versa, any program constructed by the DSL can be used in the host.
This is in contrast to deep embedding where host programs must be converted to an AST representation.
Such conversion is difficult to achieve transparently especially when the built-in functions and constructs, such as conditionals, are encountered.

Typical shallowly embedded DSLs have an overhead stemming from the additional layers of functions tied to the language syntax.
As we have shown in \autoref{ch:building_stage} however, dynamic staging can resolve all overhead coming from the embedding,
producing only the intended user code.
Moreover, unlike typical shallow embeddings, the produced code is not directly tied to the DSL syntax.
The actual computation may occur later or even in a different order, as in the chained matrix example in \autoref{ch:usage:matrix}.
It is also possible for a DSL to produce not one but two or more chains of fragment functions.
Such technique can be used to generate a multi-pass interpretation in a single parsing, e.g. where circular definitions are present.

On the other hand, our functional builders resemble the deep embedding as well:
Actions produce code objects and the code is built incrementally,
by connecting the fragment functions in a tree structure similar to an AST.

Nodes in our graph structure are self-descriptive,
i.e. their behaviour is entirely defined within the functions they contain.
As shown in \autoref{ch:usage:env} these functions may contain much more than just the core semantics.
The user may define many staging phases and incorporate additional, auxiliary computation within it.
It may be used to define custom name lookups, but also as an analysis step, optimization, or even perform custom type checking~\cite{AuxiliaryComp}.

Unlike the deep embedding however, we do not support tree transformations.
Optimization is not achieved through analysis of the program graph, but through the partial evaluation enabled by dynamic staging.
We do not claim that it is the only way programs should be optimized:
careful analysis of the code is needed to generate fast, low-level machine code.
What we claim, instead, that such low-level techniques are usually not needed at higher-level DSL definitions and generally should be avoided because of their complexity.

This paper focuses on the conceptual aspects of building code through dynamic staging.
However, it is important to state that the presented approach has been fully implemented in the DeepCPS interpreter and compiler.
This includes the examples presented here, as well as more complex DSLs, such as a language for grammar specification and for stencil code generation.
In all the cases we were able to specify domain optimization, the interpreter was able to remove any residual overhead and the compiler produced an efficient machine code.

\section{Conclusions and Future Work}
\label{sec:conclusion}

We have presented a novel approach to code building.
We use an entirely functional approach, where each builder produces a fragment function containing an arbitrary program code.
We have shown how builders are defined and argued how dynamic staging allows us to remove the overhead coming from such encapsulation.

In \autoref{ch:usage} we have given simple examples of how typical language design problems can be handled using builders and dynamic staging.
This includes domain-specific optimizations, which are also represented as additional code.
It is more natural for for a domain-expert to implement an optimizing algorithm as a program, rather than as an AST transformation.
The whole optimization algorithm can be hidden away in a single builder node, or even provided as a separate function, independently of the DSL definition.

We did not include the performance numbers for the machine code produced by our DSLs as we do not evaluate the efficiency of the DeepCPS compiler.
Such evaluation has been done in~\cite{DeepCPS}.
Note, that we are able to produce the DeepCPS code containing no unnecessary residual operations originating from the builders.
The back-end compiler receives the same input as if the code was written without the builders, and produces the same result.

\subsection*{Future Work}

The focus in this paper is only on the semantic actions of a language.
For a fully functioning DSL this is not enough -- a parser or an embedding is needed.
We did not provide any integration of our semantic actions with a parser, nor explained how to embed a DSL in DeepCPS.
Such integration is in our current focus and we have made substantial advancements in that area.
This is however not in the scope of this paper.

In \autoref{ch:usage:env} we described how dynamic staging can be used to define an auxiliary computation.
The same approach can be used to build custom type systems.
In theory, any type system can be defined as a staged computation and used in a DSL.
To our knowledge this possibility has not yet been fully explored and requires further research.
This may be particularly interesting in the context of DeepCPS as dynamic staging may bring dynamic and static typing closer together,
in a similar way as it did for shallow and deep embeddings.

We also hope to explore other uses for functional code building, not directly related to language definitions.
As each fragment can be an arbitrary large piece of code, the same technique can be used to build programs at bigger granularity.
For example, this technique may become viable for data-flow programming, and semi-automatic, profiled assembly.

\printbibliography

\end{document}